\documentclass[twocolumn,showpacs,superscriptaddress,prl,amsmath,amssymb,letterpaper]{revtex4}

\usepackage{graphicx}
\usepackage{multirow}
\usepackage{color}
\usepackage{bm}
\usepackage{ulem}
\usepackage{times}
\usepackage{amsmath,bm,amsfonts}
\usepackage{dcolumn}
\usepackage{graphicx}
\usepackage{latexsym}
\usepackage{mathrsfs} %%%%% included in 191118
\usepackage{ bbold }%%%%% included in 191121

\def\ket#1{|#1\rangle }
\def\bra#1{\langle #1 |}

\begin{document}

\title{Many-body approach to non-Hermitian physics in fermionic systems}

\author{Eunwoo \surname{Lee}}\thanks{These authors contributed equally to this work}
\affiliation{Department of Physics and Astronomy, Seoul National University, Seoul 08826, Korea}

\affiliation{Center for Correlated Electron Systems, Institute for Basic Science (IBS), Seoul 08826, Korea}

\affiliation{Center for Theoretical Physics (CTP), Seoul National University, Seoul 08826, Korea}

\author{Hyunjik \surname{Lee}}\thanks{These authors contributed equally to this work}
\affiliation{Department of Physics and Astronomy, Seoul National University, Seoul 08826, Korea}

\affiliation{Center for Correlated Electron Systems, Institute for Basic Science (IBS), Seoul 08826, Korea}

\affiliation{Center for Theoretical Physics (CTP), Seoul National University, Seoul 08826, Korea}

\author{Bohm-Jung \surname{Yang}}
\email{bjyang@snu.ac.kr}
\affiliation{Department of Physics and Astronomy, Seoul National University, Seoul 08826, Korea}

\affiliation{Center for Correlated Electron Systems, Institute for Basic Science (IBS), Seoul 08826, Korea}

\affiliation{Center for Theoretical Physics (CTP), Seoul National University, Seoul 08826, Korea}

\date{\today}

\begin{abstract}
In previous studies, the topological invariants of one-dimensional non-Hermitian systems have been defined in open boundary condition (OBC) to satisfy the bulk-boundary correspondence.
The extreme sensitivity of bulk energy spectra to boundary conditions has been attributed to the breakdown of the conventional bulk-boundary correspondence based on the topological invariants defined under periodic boundary condition (PBC).
Here we propose non-Hermitian many-body polarization as a topological invariant for 1D non-Hermitian systems defined in PBC, which satisfies the bulk-boundary correspondence.
Employing many-body methodology in the non-Hermitian Su-Schrieffer-Heeger model for fermions, we show the absence of non-Hermitian skin effect due to the Pauli exclusion principle and demonstrate the bulk-boundary correspondence using the invariant defined under PBC. 
Moreover, we show that the bulk topological invariant is quantized in the presence of chiral or generalized inversion symmetry.
Our study suggests the existence of generalized crystalline symmetries in non-Hermitian systems, which give quantized topological invariants that capture the symmetry-protected topology of non-Hermitian systems.
\end{abstract}

\pacs{}

\maketitle

%{\it Introduction.|}

%{\it Introduction.|}
%The attempts to describe open or dissipative quantum systems have led to the establishment of a new field: non-Hermitian quantum mechanics~\cite{znojil2001conservation, bender2005introduction, bender2007making, rotter2009non, diehl2011topology}.Recent studies on non-Hermitian system have demonstrated various intriguing physical properties, which are absent in Hermitian systems, such as sensitive change of bulk energy spectra, known as non-Hermitian skin effect, and characteristic complex energy spectra including exceptional surfaces or bulk Fermi-arcs~\cite{yang2019non, yoshida2019symmetry, budich2019symmetry, okugawa2019topological, moors2019disorder, carlstrom2018exceptional, zhou2019exceptional, heiss2012physics, carlstrom2019knotted, midya2018non, zhou2018observation, kozii2017non, mostafazadeh2002pseudo1, mostafazadeh2002pseudo2, mostafazadeh2002pseudo3, brody2013biorthogonal}.Especially, the topological property of non-Hermitian systems is now one of the main concerns.Bulk boundary correspondence (BBC) is the fundamental property of topological phases.In this regard, various methods have been developed to study the BBC of non-Hermitian systems with only partial success, such as generalized Brillouin Zone (GBZ)~\cite{song2019non2, yokomizo2019non, deng2019non, yao2018non}, transfer matrix~\cite{kunst2019non} and entanglement spectrum~\cite{herviou2019entanglement}.However, there are two major limitations in preceding studies of non-Hermitian systems.

{\it Introduction.}
Recent progress in the study of non-Hermitian systems, such as open systems or dissipative systems with gain and loss~\cite{znojil2001conservation, bender2005introduction, bender2007making, rotter2009non, carmichael1993quantum, diehl2011topology, malzard2015topologically, lee2014heralded, lee2014entanglement, makris2008beam, klaiman2008visualization, regensburger2012parity, ruter2010observation, lin2011unidirectional, feng2013experimental, guo2009observation, liertzer2012pump, peng2014loss, fleury2015invisible, chang2014parity, hodaei2017enhanced, hodaei2014parity, feng2014single, gao2015observation, xu2016topological, ashida2017parity, kawabata2017information, chen2017exceptional, ding2016emergence, downing2017topological, ozawa2019topological, lu2014topological, el2018non, longhi2018parity, konotop2016nonlinear, cao2015dielectric, doppler2016dynamically, lapp2019engineering, lee2019topological, ashida2018full}, has uncovered various intriguing physical phenomena that do not exist in Hermitian systems. For instance, the characteristic complex energy spectra of non-Hermitian systems are theoretically predicted to host exceptional surfaces or bulk Fermi-arcs~\cite{yang2019non, yoshida2019symmetry, budich2019symmetry, okugawa2019topological, moors2019disorder, carlstrom2018exceptional, zhou2019exceptional, heiss2012physics, carlstrom2019knotted, midya2018non, kozii2017non, mostafazadeh2002pseudo1, mostafazadeh2002pseudo2, mostafazadeh2002pseudo3, brody2013biorthogonal, lee2018tidal, li2019geometric, yoshida2018non}, which are later realized in experiments ~\cite{zhou2018observation, ding2018experimental}. Nowadays, there are growing research activities to extend the idea of topological Bloch theory developed in Hermitian systems to non-Hermitian Hamiltonians~\cite{song2019non2, yokomizo2019non, deng2019non, yao2018non, kunst2019non, herviou2019entanglement, edvardsson2019non,zirnstein2019bulk,imura2019generalized, wu2019inversion, yoshida2019non, gong2018topological}.

One central issue in the study of topological phenomena in non-Hermitian systems is to understand the bulk-boundary correspondence (BBC). In Hermitian systems, it is well-established that the bulk topological invariants defined by Bloch wave functions in periodic boundary condition (PBC) predict robust boundary states in systems under open boundary condition (OBC) 
~\cite{ryu2006entanglement, grusdt2013topological, rhim2017bulk}. Contrary to this, in non-Hermitian systems, the bulk energy spectra exhibit extreme sensitivity to boundary conditions~\cite{kunst2018biorthogonal, xiong2018does, jin2019bulk}.
For instance, in recent studies of the non-Hermitian Su-Schrieffer-Heeger (SSH) model, it was shown that the bulk eigenstates, which are extended under PBC, are exponentially localized on one-side of the finite-size system with OBC ~\cite{yao2018edge, yokomizo2019non}. This phenomenon is named the non-Hermitian skin effect in Ref.~\onlinecite{yao2018edge}, which has been extensively discussed recently~\cite{yao2018edge, song2019non1, longhi2019probing, alvarez2018topological, lee2019anatomy, borgnia2019non, zhang2019correspondence, okuma2019topological, hofmann2019reciprocal, helbig2019observation, lee2019hybrid}.
Since the energy spectra under PBC and OBC differ so drastically, there has even been a common belief that the bulk invariant defined under PBC has intrinsic limitations in explaining the BBC of non-Hermitian systems in general.
To circumvent this problem, several interesting theoretical ideas are proposed under OBC, such as generalized Bloch theory~\cite{song2019non2, yokomizo2019non, deng2019non, yao2018non}, transfer matrix approach~\cite{kunst2019non}, and entanglement spectrum analysis~\cite{herviou2019entanglement}.
The lack of proper topological invariants defined under PBC that satisfy BBC gives the impression that the BBC of non-Hermitian systems belongs to a rather special category which is distinct from that of Hermitian systems. However, is it really true that the BBC of non-Hermitian topological systems evades the theoretical framework developed to understand the topological phases in Hermitian systems? 

Here we address this important question focusing on one-dimensional (1D) fermionic non-Hermitian systems. For the non-Hermitian SSH model describing spinless fermions, we show that the non-Hermitian skin effect, observed in a single-particle approach, does not appear in the many-body approach due to the Pauli exclusion principle~\cite{mu2019emergent}. We have also found that at half-filling, topologically trivial and non-trivial phases display the same charge density distribution in systems with OBC. When one extra electron or hole is added, however, the additional charge is exponentially localized near the edges when the system is topologically non-trivial, whereas it spreads over the entire system when the system is topologically trivial.

Moreover, we have identified a bulk topological invariant defined under PBC: the non-Hermitian many-body polarization, which correctly describes the BBC and gives the same bulk critical points for topological phase transitions as those predicted under OBC~\cite{yao2018edge}. We find that the many-body polarization defined under PBC is quantized in the presence of chiral or generalized inversion symmetry. This clearly shows that, in non-Hermitian systems, one can define bulk topological invariants under PBC, which are quantized due to generalized crystalline symmetries and correctly predict the associated boundary states, as in the case of conventional fermionic Hermitian systems.
Finally, we propose the non-Hermitian version of the edge entanglement entropy that can be used to detect the edge degeneracy in 1D non-Hermitian systems.

%%%%%%%%%%%%%%%%%%%%%%%%%%%%%%%%%%%%%%%%%%%%%%%%%%%%%%%%%%%%%%%%%
\begin{figure}[t]
\centering
\includegraphics[width=8.5cm]{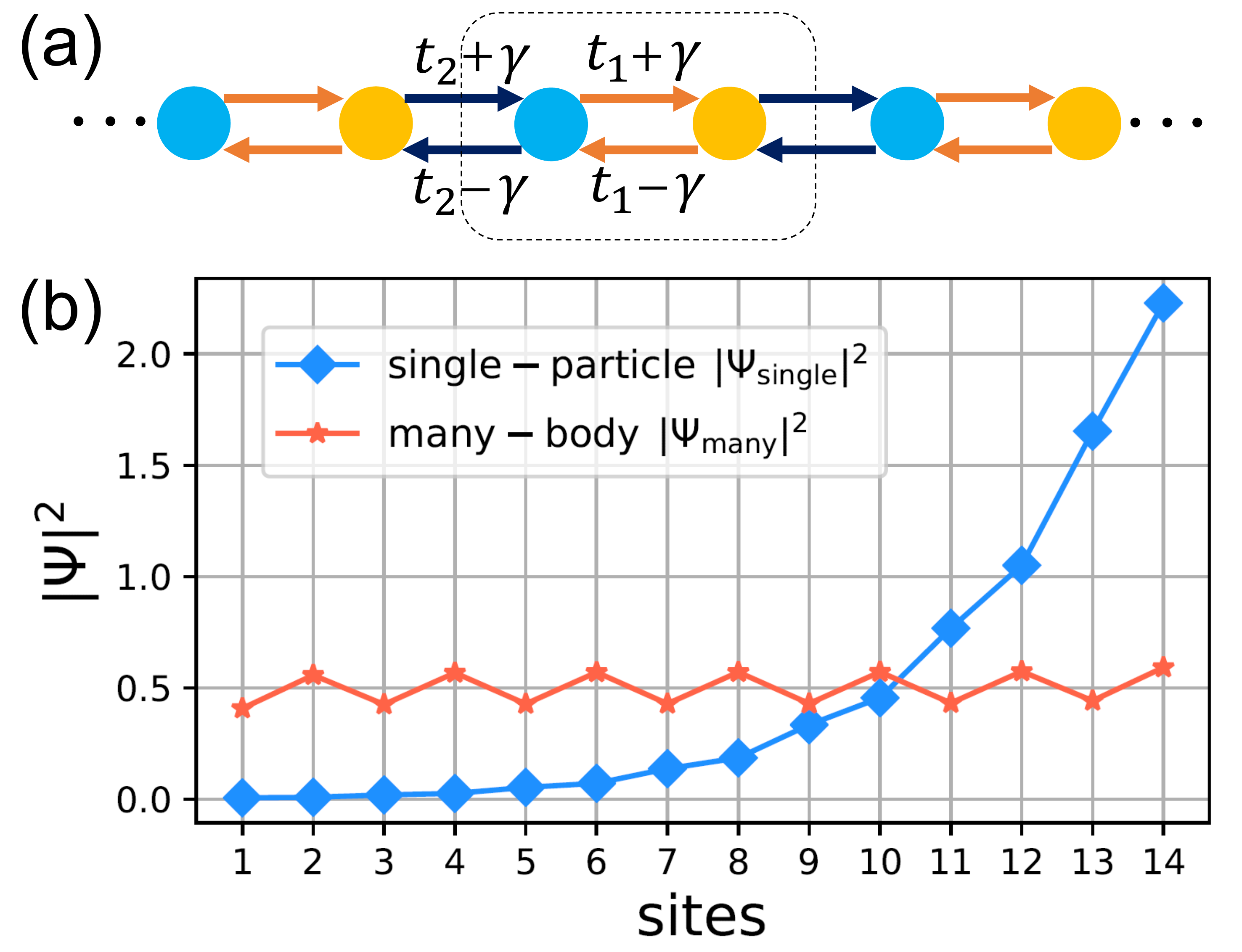}
\caption{
(a) Schematic figure describing the non-Hermitian SSH model for spinless fermions. The dotted box denotes the unit cell. $t_1$ ($t_2$) indicates the intracell (intercell) hopping while $\gamma$ denotes the asymmetric hopping term that makes the system non-Hermitian.
(b) Distribution of the particle density $|\Psi(i)|^{2}$ from single-particle wave functions (square dots) and many-body wave functions (star-shaped dots) obtained by solving a finite-size chain with 14 sites under OBC. We choose the model parameters $t_1=2, t_2=1, \gamma=0.3$ which correspond to a trivial insulator. 
To obtain $|\Psi(i)|^{2}$ from single-particle wave functions, the contributions from all states below the gap are added.
%Almost all wave functions are exponentially localized, manifesting the non-Hermitian skin effect.
%Meanwhile, the particle density $|\Psi_{RR}|^{2}$ from many-body wave function is almost uniform with small oscillation. Due to the Pauli exclusion principle, non-Hermitian skin effect is absent when many-body wave functions are used.
} \label{fig:SSH}
\end{figure}
%%%%%%%%%%%%%%%%%%%%%%%%%%%%%%%%%%%%%%%%%%%%%%%%%%%%%%%%%%%%%%%

{\it Model.} 
The non-Hermitian SSH model Hamiltonian $\hat{H}_{\rm SSH}$ is composed of two parts as $\hat{H}_{\rm SSH}=\hat{H}_{0}+\hat{H}_{\rm NH}$ in which
%%%% equations 1 & 2
\begin{align}
\hat{H}_{0}&=\sum_{i}\left\{\left[J-\Delta J(-1)^{i}\right]\hat{c}_{i}^{\dagger}\hat{c}_{i+1}+\rm{H.c}.\right\},\\
\hat{H}_{\rm NH}&=\sum_{i}\gamma(\hat{c}_{i+1}^{\dagger}\hat{c}_{i}-\hat{c}_{i}^{\dagger}\hat{c}_{i+1}),
\end{align}
%%%%
where $\hat{c}_i$ ($\hat{c}_i^{\dagger}$) denotes the electron annihilation (creation) operator at the $i$-th site.
$\hat{H}_{0}$ indicates the Hermitian SSH Hamiltonian with the intracell (intercell) hopping amplitude $t_{1}=J+\Delta J$ ($t_{2}=J-\Delta J$) while $\hat{H}_{\rm NH}$ denotes the non-Hermitian part describing asymmetric hopping processes [see Fig.~\ref{fig:SSH} (a)].
The model is equivalent to a Creutz-ladder-like system with gain and loss, which is experimentally realizable in an ultracold fermionic system~\cite{yokomizo2019non,SM}.

Diagonalizing the Hamiltonian $\hat{H}_{\rm SSH}$ under OBC and using the corresponding single-particle eigenvector $\ket{\psi_n}$  with the band index $n$, one can obtain the particle density at the $i$th site $|\Psi_{\rm single}(i)|^2\equiv\sum_{n\in \textrm{occ}}\bra{\psi_n}\hat{c}_{i}^{\dagger}\hat{c}_{i}\ket{\psi_{n}}$. Here we have added the contribution from all the single-particle wave functions below the Fermi level, as
plotted in Fig.~\ref{fig:SSH} (b) with square dots.
One can observe the exponential localization of the particle density on the right edge, manifesting the non-Hermitian skin effect.
However, this accumulation obviously violates the Pauli exclusion principle, which should be corrected to obtain physically meaningful results~\cite{mu2019emergent}.

To take into account of the Fermi statistics, we take the many-body approach using Fock basis states.   
For a system with $L$ lattice sites filled with $N$ electrons, the number of allowed Fock basis states is given by $\binom{L}{N}$.
For example, a four-site system filled with two electrons has six Fock basis states: $\ket{1100}, \ket{1010}, \ket{1001}, \ket{0110}, \ket{0101}, \ket{0011}$ where $0$ and $1$ indicate the number of electron at each site. 
The Hamiltonian acts on Fock bases as follows: $\hat{c}_{3}^\dagger \hat{c}_{2}\ket{1100}=\ket{1010}, ~\hat{c}_{3}^\dagger \hat{c}_{2}\ket{0110}=0$, and so on.
Since a single-particle Hamiltonian is of the form $\hat{H}=\sum_{i,j}\mathrm{H}_{ij}\hat{c}_{i}^{\dagger}\hat{c}_{j}$ where $\mathrm{H}_{ij}$ denotes its $(i,j)$-th element, the $(\alpha,\beta)$-th element of the Hamiltonian in Fock space $\mathrm{H^{F}}_{\alpha\beta}$ can be obtained as $\mathrm{H^{F}}_{\alpha\beta}=\pm\mathrm{H}_{ij}$ when $\bra{\alpha}\hat{c}_{i}^{\dagger}\hat{c}_{j}\ket{\beta}=\pm1$ for Fock bases $\ket{\alpha}, \ket{\beta}$, while $\mathrm{H^{F}}_{\alpha\beta}=0$ otherwise.
$\mathrm{H^{F}}$ is diagonalized using the exact diagonalization method.

In Fig.~\ref{fig:SSH} (b), we plot the particle density distribution $|\Psi_{\rm many}(i)|^2\equiv\bra{\Psi_{R}^{G}}\hat{c}_{i}^{\dagger}\hat{c}_{i}\ket{\Psi_{R}^{G}}$ with star-shaped dots, where $\ket{\Psi_{R}^{G}}$ is the many-body ground state wave function that satisfies $\mathrm{H}^{\rm F}\ket{\Psi_{R}^{G}}=E\ket{\Psi_{R}^{G}}$. 
The sum of the wavefunction density is normalized to $N$, which is the total particle number of the system.
One can observe only mild slanting of particle densities instead of the exponential localization~\cite{SM}.
The absence of non-Hermitian skin effect in fermionic systems prompts a question: what happens to BBC when many-body formalism is used?

{\it Biorthogonal formulation.|} 
To describe the topological property of the many-body wave functions, we adopt the formulation of biorthogonal quantum mechanics. In general, a diagonalizable non-Hermitian matrix $\rm H$ can have different left and right eigenvectors $\ket{\Psi_{L}}, \ket{\Psi_{R}}$ that satisfy~\cite{brody2013biorthogonal}
\begin{align}
\mathrm{H}\ket{\Psi_{R,n}}=E_{n}\ket{\Psi_{R,n}}, \;\; \mathrm{H}^{\dagger}\ket{\Psi_{L,n}}=E_{n}^{*}\ket{\Psi_{L,n}}.
\end{align}
The sets of eigenvectors can be chosen to satisfy the biorthonormality condition $\bra{\Psi_{L,n}}\Psi_{R,m}\rangle=\delta_{nm}$
, which ensures that the transition amplitude $|\bra{\Psi_{L,n}}\Psi_{R,m}\rangle|^{2}$ between two states with different energy eigenvalues is zero.
In this paper, we focus on the Hamiltonian matrices with biorthogonality.

The biorthogonal bases naturally construct the identity operator as
$\hat{\mathbb{1}}=\sum_{n}\ket{\Psi_{R,n}}\bra{\Psi_{L,n}}$.
The expectation value of the observable in a pure right state $\ket{\Psi_{R}}=\sum_{n}C_{n}\ket{\Psi_{R,n}}$ is expressed as 
\begin{align}
\langle\hat{O}\rangle=\langle\Psi_{L}|\hat{O}|\Psi_{R}\rangle,
\end{align}
where $\ket{\Psi_{L}}=\sum_{n}C_{n}\ket{\Psi_{L,n}}$, such that $\bra{\Psi_{L}}\Psi_{R}\rangle=\sum_{n}C_{n}^{*}C_n=1$.
The corresponding particle density at the site $i$ is given by $|\Psi_{LR}(i)|^2\equiv\bra{\Psi_{L}}\hat{c}_{i}^{\dagger}\hat{c}_{i}\ket{\Psi_{R}}$. Let us note that $|\Psi_{LR}(i)|^2$ is different from the conventional particle density, which is relevant to $|\Psi_{RR}(i)|^2\equiv\bra{\Psi_{R}}\hat{c}_{i}^{\dagger}\hat{c}_{i}\ket{\Psi_{R}}/\langle\Psi_{R}|\Psi_{R}\rangle$ in terms of biorthogonal formulation.

%%%%%%%%%%%%%%%%%%%%%%%%%%%%%%%%%%%%%%%%%%%%%%%%%%%%%%%%%%%%%%%%%%%
\begin{figure}[t]
\centering
\includegraphics[width=8.5 cm]{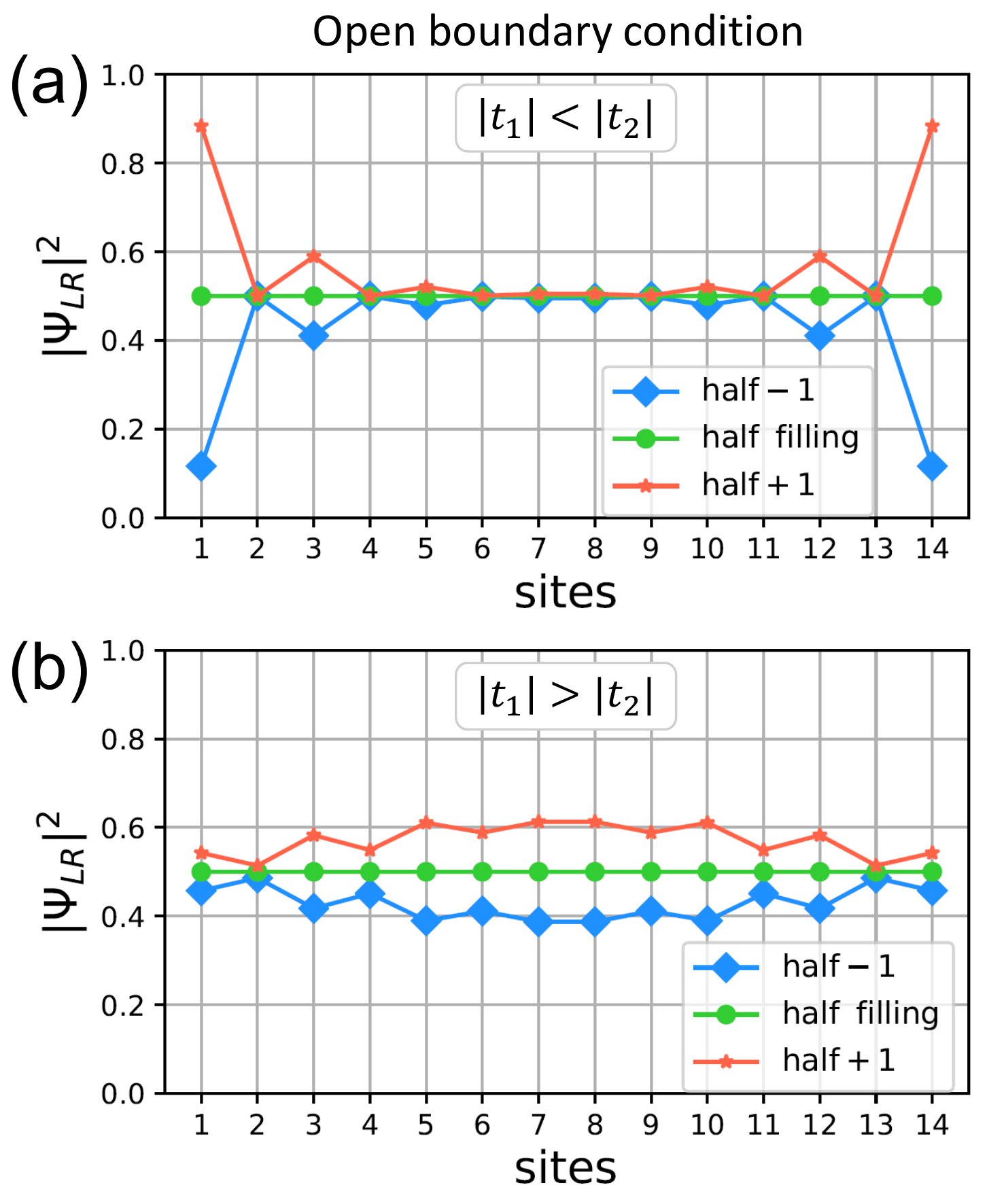}
\caption{
Distribution of $|\Psi_{LR}(i)|^{2}$ for a non-Hermitian SSH chain with 14 sites under OBC. (a) Topological insulator where $t_1=1,~t_2=2,~\gamma=0.3$.
At half-filling, $|\Psi_{LR}|^{2}$ is uniformly distributed.
Adding (subtracting) an electron, the extra electron density is accumulated (depleted) at the edges, which shows the existence of topological zero modes localized at the edges.
The zero-mode is always observed when $|t_1|<|t_2|$.
(b) Trivial insulator where $t_1=2,~t_2=1,~ \gamma=0.3$.
At half-filling, the electron density is uniformly distributed as in the case of the topological insulator.
Adding (subtracting) an electron, the extra electron density accumulated (depleted) spreads throughout the bulk, showing the absence of localized zero-modes.
} \label{fig:psi_RL}
\end{figure}
%%%%%%%%%%%%%%%%%%%%%%%%%%%%%%%%%%%%%%%%%%%%%%%%%%%%%%%%%%%%%%%%%%%

{\it Topological zero-modes.|}
One of the most prominent characteristics of the SSH model that distinguishes the topologically trivial and non-trivial phases is the existence of zero-modes at the edge.
To confirm the existence of the zero-modes in the non-Hermitian SSH model, $|\Psi_{LR}(i)|^{2}$ is computed by using the ground state many-body wave function $\{\ket{\Psi_{L}^G},\ket{\Psi_{R}^G}\}$ as shown in Fig.~\ref{fig:psi_RL}.
At half-filling, the distribution of $|\Psi_{LR}(i)|^{2}$ is uniform in both trivial and topological phases. However, when one electron or hole is added, the extra charge is localized at the edges in topological phases. In contrast, in trivial phases, the extra charge is spread over the entire system.
The localized edge modes are always present (absent) when $|t_{1}|<|t_{2}|$ ($|t_{1}|>|t_{2}|$)~\cite{SM}.
Moreover, when $|t_1|<|t_2|$, the ground state energy does not change when one electron is added or subtracted. 
This means that the states localized at the edges are indeed zero-energy edge modes.
When $|t_2|<|t_1|$, on the other hand, adding (subtracting) an electron increases (decreases) the energy as much as the energy gap.
Namely, the topological phase transition between the trivial and topological insulators occurs at $|t_2|=|t_1|$. The location of the critical point is the same as that from the generalized Brillouin zone approach under OBC~\cite{yao2018edge}.
Note that the entire energy spectrum of our model is real in OBC when $|\gamma|<|t_{1}|, |t_{2}|$~\cite{SM}.

\begin{figure}[h]
\centering
\includegraphics[width=8.5 cm]{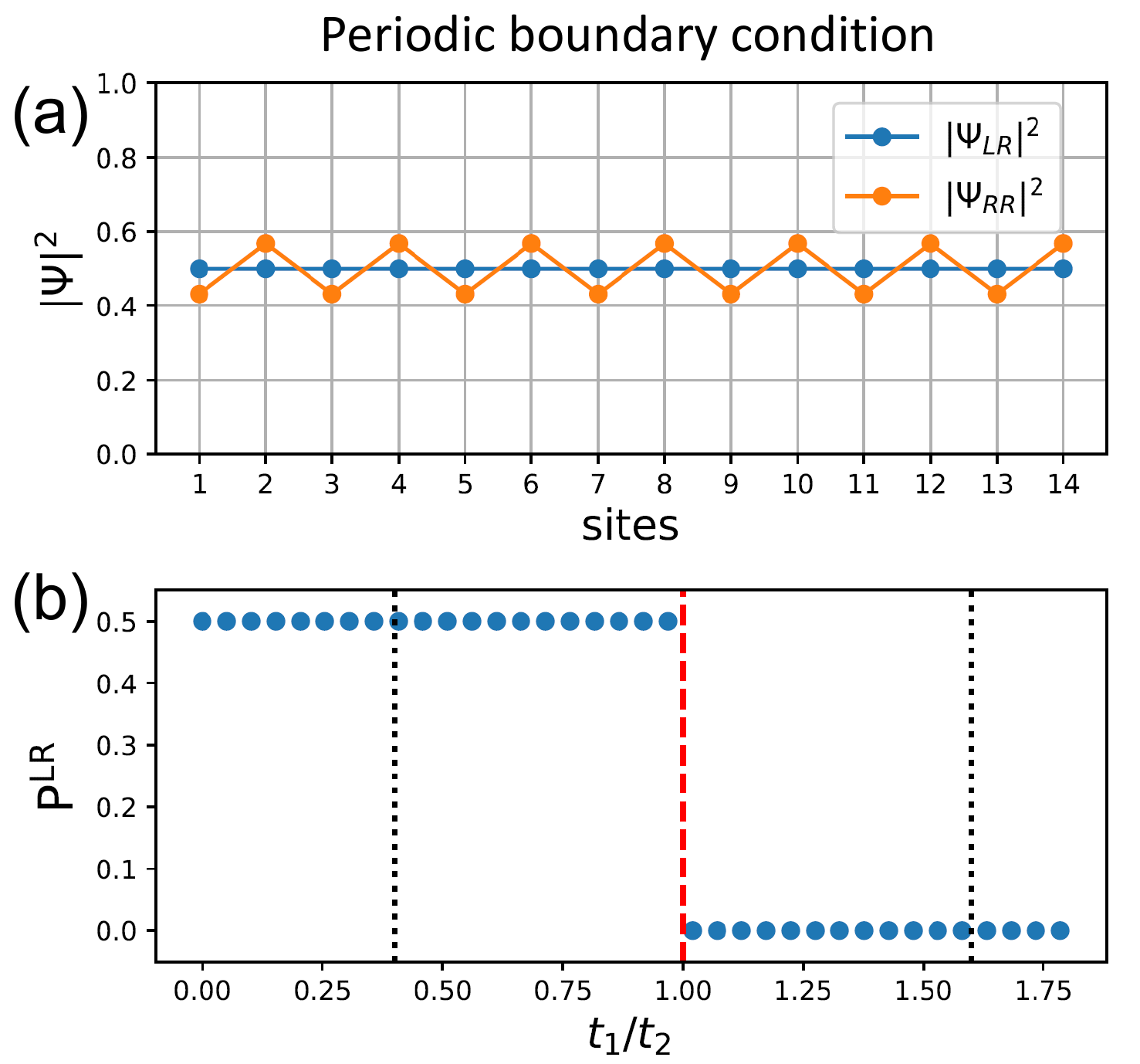}
\caption{
(a) Distribution of the particle densities $|\Psi_{LR}(i)|^{2}$ and $|\Psi_{RR}(i)|^{2}$ at half-filling, for a non-Hermitian SSH model with 14 sites under PBC.
We use the model parameters $t_1=2,~t_2=1,~\gamma=0.3$ that correspond to a trivial phase.
$|\Psi_{LR}|^{2}$ is uniform whereas $|\Psi_{RR}|^{2}$ has a sawtooth shape. 
(b) Non-Hermitian many-body polarization $P^{LR}$ as a function of $t_1$ with $t_2=1,~\gamma=0.3$. When $0<t_1<t_2$ ($t_1>t_2>0$), $P^{LR}=1/2$ ($P^{LR}=0$).
The critical point $t_1=t_2$ is consistent with the numerical study of the zero modes under OBC. Thus, the BBC can be described by using $P^{LR}$.
The black dotted vertical lines indicate the locations of gap-closing points obtained by the Bloch Hamiltonian, which cannot explain the BBC.
} \label{fig:invariant}
\end{figure}

Let us now obtain the critical points obtained under PBC. Naive application of the Bloch theory in PBC predicts the closing of bulk gaps at $t_2=t_1\pm 2\gamma$ or $t_2=-t_1$, which is not consistent with the critical point where the zero-modes appear or disappear~\cite{lee2016anomalous}.
Such a discrepancy occurs because the energy eigenvalues obtained in PBC are not real, but generally complex.
Interestingly, in our model, the many-body ground state energy of $H^{F}$ in PBC is real and thus physically stable when $|\gamma|<|t_{1}|, |t_{2}|$ and $N=L/2$ is an odd integer~\cite{SM}. The distribution of the particle density of the system in PBC is shown in Fig.~\ref{fig:invariant}(a) obtained by using the many-body ground state.

{\it Non-Hermitian polarization.|}
In order to discuss the topological phase in PBC, we need to define a bulk topological invariant.
In Hermitian systems, the many-body bulk polarization is a well-defined 1D topological invariant, whose definition under PBC is given by~\cite{resta1992theory}
\begin{align}
P\equiv\lim_{N\rightarrow\infty}\frac{1}{2\pi}\text{Im}\ln\langle\Psi^G|e^{i(2\pi/N)\hat{X}}|\Psi^G\rangle \mod 1,
\end{align}
where $|\Psi^G\rangle$ is the many-body ground state and $N$ is the number of unit cells. Here $\hat{X}$ denotes the summation of position operators for all atoms.
$P$ is quantized in the presence of either inversion or chiral symmetry.
The Hamiltonian $\hat{H}$ invariant under the chiral $\hat{S}$ or inversion $\hat{I}$ symmetry satisfies
\begin{align}
\hat{S}\hat{H}\hat{S}^{-1}=\hat{H},\quad
\hat{I}\hat{H}\hat{I}^{-1}=\hat{H}, 
\end{align}
where $\hat{S}\hat{c}_{i}\hat{S}^{-1}=(-1)^{i}\hat{c}_{i}^{\dagger}$, $\hat{S}i\hat{S}=-i$ and $\hat{I}\hat{c}_{i}\hat{I}^{-1}=\hat{c}_{L+1-i}$, $\hat{I}i\hat{I}^{-1}=i$~\cite{chiu2016classification}.
In terms of the corresponding matrix representation $S$ and $I$, the invariance of the Hamiltonian matrix $H$ becomes
\begin{align}
S^{-1}HS=-H,\quad I^{-1}HI=H.
\end{align}
Under inversion symmetry, the polarization satisfies $P \equiv -P \mod 1$, so that it is quantized into either 0 or $1/2 \mod 1$.
Also, chiral symmetry imposes $P_{\text{occ}}=P_{\text{unocc}}$ with $P_{\text{occ}}+P_{\text{unocc}}=0 \mod 1$, which lead to the polarization quantization: $P_{\text{occ}}=0$ or $1/2 \mod 1$. Here $P_{\text{occ}}$ ($P_{\text{unocc}}$) denotes the polarization of occupied (unoccupied) states.

We extend the idea of the many-body bulk polarization, which has been used in Hermitian systems only, to non-Hermitian systems by defining the non-Hermitian many-body bulk polarization $P^{LR}$ as
\begin{align}
P^{LR}\equiv\lim_{N\rightarrow\infty}\frac{1}{2\pi}\text{Im}\ln\langle\Psi^{G}_{L}|e^{i(2\pi/N)\hat{X}}|\Psi^{G}_{R}\rangle \mod 1,
\end{align}
where $|\Psi^{G}_{R}\rangle (\ket{\Psi^{G}_{L}})$ is the right (left) many-body ground state. 
Here, we introduce chiral and generalized inversion symmetries for non-Hermitian systems, which quantize $P^{LR}$, as follows
\begin{align}
\hat{S}\hat{H}\hat{S}^{-1}=\hat{H}^{\dagger},\quad
\hat{I}\hat{H}\hat{I}^{-1}=\hat{H}^{\dagger}.
\end{align}
In terms of the corresponding matrix representation, we have
\begin{align}
S^{-1}HS=-H, \quad
I^{-1}HI=H^{\dagger}.
\end{align}
Note that $\hat{H}_{\rm SSH}$ has both chiral and generalized inversion symmetry.
One can also check the existence of these symmetries in Fock space representation as well.
Under generalized inversion symmetry, $P^{LR}\equiv-P^{LR} \mod 1$, and thus $P^{LR}$ is quantized into $0$ or $1/2$.
Likewise, under chiral symmetry, $P^{LR}_{\text{occ}}=P^{LR}_{\text{unocc}}$ results in the quantization of $P^{LR}_{\text{occ}}$ with the condition $P^{LR}_{\text{occ}}+P^{LR}_{\text{unocc}}=0\mod 1$~\cite{SM}. Therefore in the presence of either chiral or generalized inversion symmetry, the non-Hermitian many-body polarization is quantized into either $0$ or $1/2$.

As shown in Fig.~\ref{fig:invariant} (b), $P^{LR}$ defined under PBC is $1/2$ when $|t_1|<|t_2|$ whereas it is $0$ when $|t_1|>|t_2|$.
The phase transition occurs at the critical point $|t_1|=|t_2|$, which is consistent with the numerical study of the zero-modes in OBC discussed before. In fact, one can understand the reason why the critical point is located at $|t_1|=|t_2|$ as follows.
Switching the role of $t_1$ and $t_2$ is equivalent to translating the system by a half lattice constant. Then the $i$th site moves to the $(i+1)$th site. Thus, $P^{LR}$ with $t_1=\alpha, t_2=\beta$ is different by $1/2$ from $P^{LR}$ with $t_1=\beta, t_2=\alpha$~\cite{SM}. The location of the critical point agrees with the numerical and analytical results obtained under OBC~\cite{SM} indicating that BBC is satisfied when $P^{LR}$ defined under PBC is considered.
This is also confirmed in another model as well~\cite{SM}.

%Furthermore,  to find out whether both chiral and inversion symmetries protect the invariant, we i) break $\hat{S}$ but keep $\hat{I}$ by introducing next-nereast hopping interaction and ii) break $\hat{I}$ but keep $\hat{S}$ by enlarging unit-cell twice and modifying nearest interaction from $\{t_1, t_2, t_1, t_2\}$ to $\{t_1,t_2,t_1+\delta, t_2+\delta\}$. The polarization is quantized in both cases.

%The bulk-boundary correspondence is also confirmed by the another type of SSH Hamiltonian introduced in ref~\cite{yao2018edge}.
%Here, the model Hamiltonian $\hat{H}=\hat{H}_0+\hat{H}_{\text{NH}}$:
%\begin{align}
%\hat{H}_{0}&=\sum_{i}\{(J-\Delta J(-1)^{i}\hat{c}_{i}^{\dagger}\hat{c}_{i+1}+h.c.)\}, \nonumber\\
%\hat{H}_{\rm NH}&=\sum_{i}\gamma(\hat{c}_{2i}^{\dagger}\hat{c}_{2i-1}-\hat{c}_{2i-1}^{\dagger}\hat{c}_{2i}),
%\end{align}
%where the asymmetric hopping interaction exists only for intracell electrons.
%In this model, the critical point obtained in OBC is located at $t_1=\sqrt{t_2^2+\gamma^2}$~\cite{yao2018edge}. The result agrees with the critical point of the bulk-invariant as well.

%%%%%%%%%%%%%%%%%%%%%%%%%%%%%%%%%%%%%%%%%%%%%%%%%%%%%%%%%%%%%%%%%%%
%%%%%%%%%%%%%%%%%%%%%%%%%%%%%%%%%%%%%%%%%%%%%%%%%%%%%%%%%%%%%%%%%%%

{\it Edge entanglement entropy.}
Another way to determine the topological property of many-body systems is to calculate the edge entanglement entropy, which is known to be useful in detecting the edge degeneracy~\cite{wang2015detecting, ryu2006entanglement, fidkowski2010entanglement, grover2013entanglement}.
In particular, when there are two zero modes localized at the edges of a 1D topological isulator, edge entanglement entropy is quantized to $\ln{2}$~\cite{wang2015detecting}.

To define the entanglement entropy, we consider a system that is divided into two subsystems $A$ and $B$.
In Hermitian systems, the R\'{e}nyi entropy $S_{\alpha}$ of order $\alpha$ is defined as
\begin{align}
S_{\alpha}=\frac{1}{1-\alpha}\ln \mathrm{Tr}[(\rho_{A})^{\alpha}],
\end{align}
where $\rho_{A}$ is the reduced density matrix for the subsystem $A$ and $\alpha\geq0,$ $\alpha\neq1$.
%Let us note that in $\alpha\rightarrow 1$ limit, the entropy becomes identical to von Neumann entropy $S_{1}=-\text{Tr}\rho\ln\rho$.
Similar to the way of defining non-Hermitian many-body polarization, we introduce R\'{e}nyi entropy $S_{\alpha}$ for non-Hermitian systems as
\begin{align}
S^{LR}_{\alpha}\equiv\frac{1}{1-\alpha}\ln \mathrm{Tr}[(\rho_{A}^{RL})^{\alpha}],
\end{align}
where $\rho^{RL}=|\Psi_{R}\rangle\langle\Psi_{L}|$.

The edge entanglement entropy $S^{LR}_{\alpha, \text{edge}}$ is defined as 
\begin{align}
S^{LR}_{\alpha, \text{edge}}\equiv S^{LR}_{\alpha, \text{OBC}}-\frac{1}{2}S^{LR}_{\alpha,\text{PBC}},
\end{align}
where $S^{LR}_{\alpha, \text{OBC}}$ $(S^{LR}_{\alpha, \text{PBC}})$ is calculated under OBC (PBC). 
Since the entanglement entropy follows the area law, the leading term of the entanglement entropy $S^{LR}_{\alpha, \text{PBC}}$ is about twice larger than that of $S^{LR}_{\alpha, \text{OBC}}$.
Thus, $\frac{1}{2}S^{LR}_{\alpha,\text{PBC}}$ is subtracted to cancel out the leading terms, and what remains in $S^{LR}_{\alpha, \text{edge}}$ is the sub-leading term that detects degenerate edge states in OBC that is quantized to $0$ or $\ln 2$ in thermodynamic limit.

\begin{figure}[h]
\centering
\includegraphics[width=8.5 cm]{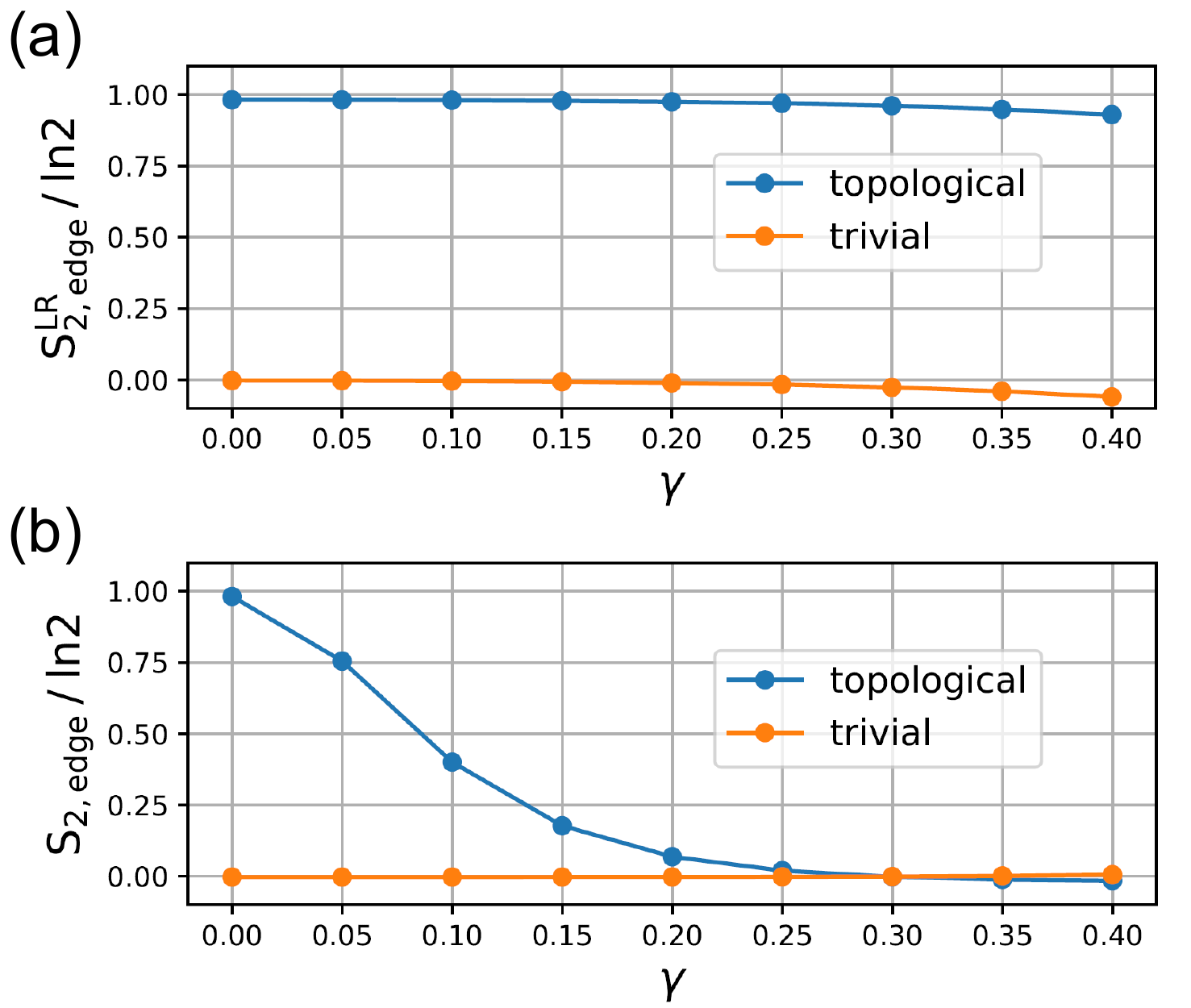}
\caption{
(a) Non-Hermitian edge entanglement entropy $S^{LR}_{2, \text{edge}}$ which detects edge degeneracy at two edges. 
A finite-size non-Hermitian SSH chain with 14 sites is considered.
$S^{LR}_{2, \text{edge}}=\ln 2$ if the ground state is topological whereas $S^{LR}_{2, \text{edge}}=0$ if the ground state is trivial. 
The deviation from the quantized values at large $\gamma$ is due to the finite-size effect, which becomes smaller as the length of the system increases.
(b) The conventional edge entanglement entropy calculated using only right eigenstates. The entropy of the trivial phase is consistently zero, whereas the topological phase's entropy plummets as the asymmetric hopping amplitude increases.
} \label{fig:EEE}
\end{figure}

As shown in Fig.~\ref{fig:EEE}(a), $S^{LR}_{2, \text{edge}}$ is ln2 for a topological phase and 0 for a trivial phase, where we choose $\alpha=2$ for the convenience of calculation.
Meanwhile, if we utilize the conventional entanglement entropy $S_{2,\text{edge}}=S_{2, \text{OBC}}-\frac{1}{2}S_{2,\text{PBC}}$, the entropy plummets as the asymmetric hopping amplitude increases [see Fig.~\ref{fig:EEE}(b)]. Thus, the topological propertiy of the non-Hermitian system cannot be correctly captured unless we use the biorthogonal formalism.

%%%%%%%%%%%%%%%%%%%%%%%%%%%%%%%%%%%%%%%%%%%%%%%%%%%%%%%%%%%
%%%%%%%%%%%%%%%%%%%%%%%%%%%%%%%%%%%%%%%%%%%%%%%%%%%%%%%%%%%%%%%%%%%

{\it Discussion.}
Our theoretical approach based on many-body wave functions suggests that careful consideration of Fermi statistics leads to the recovery of the conventional BBC even in non-Hermitian systems. Since the non-Hermitian skin effect has been observed in various non-Hermitian Hamiltonians in different dimensions, and the resulting exponential accumulation of charge densities always violates the Pauli exclusion principle, we believe that our theoretical results are valid in general non-Hermitian fermionic systems. Extending the many-body approach to higher dimensional non-Hermitian systems is definitely one important direction for future research.

Moreover, the identification of the non-Hermitian many-body polarization that is quantized under generalized inversion symmetry clearly shows the interplay between the crystalline symmetries and non-Hermitian topology.
This indicates the existence of symmetry-protected topological phases even in non-Hermitian systems. We believe that this discovery will open a new avenue to search topological non-Hermitian systems protected by crystalline symmetries.
Finally, since the topological invariant is defined by using many-body formulation, our work can be easily extended to interacting non-Hermitian systems as well.

\begin{acknowledgments}
{\it Acknowledgments.}
We thank Seunghun Lee, Yoonseok Hwang, and Junyeong Ahn for useful discussions.
E.L., H.L., and B.-J.Y. were supported by the Institute for Basic Science in Korea (Grant No. IBS-R009-D1) and Basic Science Research Program through the National Research Foundation of Korea (NRF) (Grant No. 0426-20190008), and  the POSCO Science Fellowship of POSCO TJ Park Foundation (No. 0426-20180002).
This work was supported in part by the U.S. Army Research Office under Grant No. W911NF-18-1-0137. 
\end{acknowledgments}

%\bibliography{references}

%\documentclass[aps,prl,preprint,superscriptaddress,nopacs]{revtex4}\usepackage{graphicx}
%\documentclass[aps,prb,twocolumn,groupedaddress,citeautoscript,nopacs]{revtex4}
%\documentclass[aps,prb,preprint,groupedaddress,citeautoscript,nopacs]{revtex4}

\newpage

\section{Supplemental Materials for Many-body approach to non-Hermitian physics in fermionic systems}

\date{\today}

\maketitle

\tableofcontents

\setcounter{section}{0}
\setcounter{figure}{0}
\setcounter{equation}{0}

\renewcommand{\thefigure}{S\arabic{figure}}
\renewcommand{\theequation}{S\arabic{equation}}
\renewcommand{\thesection}{SM \arabic{section}}

\section{Absence of Non-Hermitian skin effect}
\begin{figure}[t]
\centering
\includegraphics[width=8.5 cm]{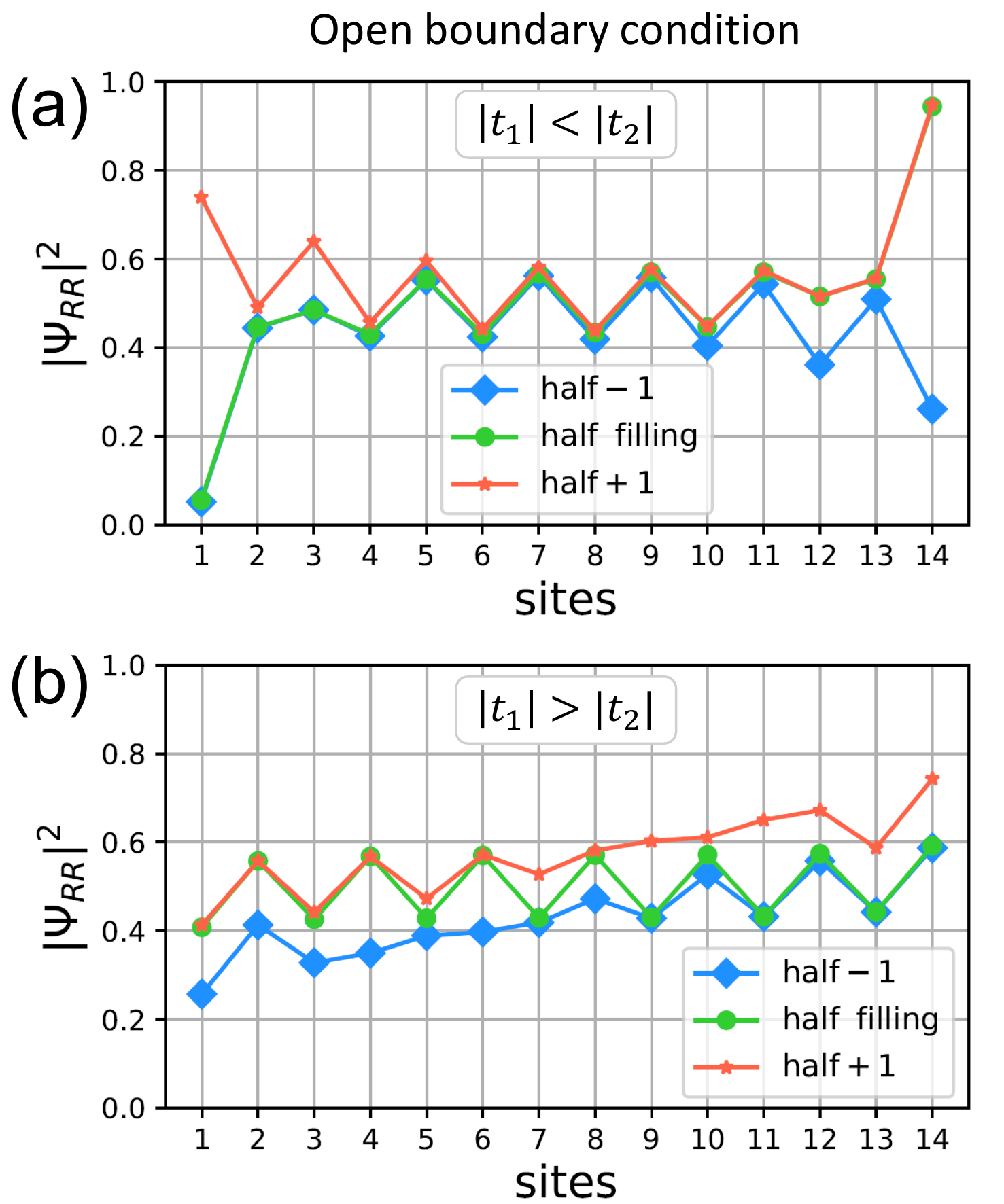}
\caption{
(Color Online) (a)$|\Psi_{RR}|^{2}$ distribution for a topological phase($t_1=1, t_2=2, \gamma=0.3$) system with 14 sites under OBC. 
At half-filling, $|\Psi_{RR}|^{2}$ is accumulated at one corner and depleted at the other corner.
Due to the Pauli exclusion principle, the maximum number at each site is restricted to $1$ and thus the non-Hermitian skin effect is absent. Topological zero-mode appears at the corners when adding or subtracting an electron.
(b)$|\Psi_{RR}|^{2}$ distribution for a trivial phase($t_1=2, t_2=1, \gamma=0.3$) system with 14 sites under OBC. 
At half-filling, the density is almost uniform with small oscillation.
Adding or subtracting an electron, energy of the ground state changes and $|\Psi_{RR}|^{2}$ is accumulated throughout the bulk.
} \label{fig:psi_RR}
\end{figure}

If we define wavefunction density as $|\Psi_{RR}(i)|^2\equiv\bra{\Psi_{R}}\hat{c}_{i}^{\dagger}\hat{c}_{i}\ket{\Psi_{R}}/\langle\Psi_{R}|\Psi_{R}\rangle$, the single-particle wavefunctions are localized at the edges exponentially. 
However, in many-body methodology, which allows at most 1 electron per site, the exponentially occupied states are not allowed due to the Pauli exclusion principle.
Fig.~\ref{fig:psi_RR} describes the many-body $|\Psi_{RR}|^{2}$ distribution for topological and trivial phases at half-filling.
For topological phase, $|\Psi_{RR}|^{2}$ is accumulated at one corner and depleted at the other corner (Fig.~\ref{fig:psi_RR} (a)).
Adding (subtracting) an electron, the density is accumulated (subtracted) at the edge, which shows the existence of the topological edge-modes.
For trivial phase, $|\Psi_{RR}|^{2}$ is almost uniform with small oscillation (Fig.~\ref{fig:psi_RR} (b)).
Adding (subtracting) an electron, the density is accumulated(subtracted) throughout the bulk, showing the absence of topological edge-modes.
Away from half-filling, it is found that $|\Psi_{LR}|^2$ is always symmetric whereas $|\Psi_{RR}|^2$ is accumulated at one edge.

%%%%%%%%%%%%%%%%%%%%%%%%%%%%%%%%%%%%%%%%%%%%%%%%%%%%%%%%%%%%%%%%%%%
%%%%%%%%%%%%%%%%%%%%%%%%%%%%%%%%%%%%%%%%%%%%%%%%%%%%%%%%%%%%%%%%%%%

\section{Energy spectrum and topological phase under OBC}

We prove that the entire energy spectrum of our model is real in OBC when $|\gamma|<|t_{1}|, |t_{2}|$.
Without loss of generality, let us assume $t_1,t_2>0$.
For later convenience, we define parameters $\alpha\equiv\sqrt{|t_{1}+\gamma|/|t_{1}-\gamma|}$, $\beta\equiv\alpha\times\sqrt{|t_{2}+\gamma|/|t_{2}-\gamma|}$, and a diagonal matrix $\mathrm{R}\equiv diag(1,\alpha,\beta,\alpha\beta,\dots,\beta^{n}, \alpha\beta^{n}, \dots)$.
One can easily check that 
\begin{align}
\rm H^{'}\equiv R^{-1}HR=H^{'\dagger}.
\end{align}
Notice that $\rm H^{'}$ is a Hamiltonian for Hermitian SSH model with intra(inter)-cellular hopping $\sqrt{t_{1}^{2}-\gamma^{2}}(\sqrt{t_{2}^{2}-\gamma^{2}})$.
This shows that the system is topological when $t_1<t_2$ and trivial when $t_1>t_2$.

When Hamiltonian $\hat{H}$ is transcribed into Fock space, we denote the Hamiltonian matrix in Fock basis as $\rm{H}^{F}$.
Since $\rm H^{'}$ is Hermitian, the transcribed matrix in Fock basis $\rm H^{'F}$ is also Hermitian.
Interpreting $\rm H^{'}$ as an operator $\hat{H}^{'}=\sum_{i,j}\mathrm{R}_{ii}^{-1}\mathrm{H}_{ij}\mathrm{R}_{jj}\hat{c}_{i}^{\dagger}\hat{c}_{j}=\sum_{i,j}\mathrm{H}_{ij}^{'}\hat{c}_{i}^{\dagger}\hat{c}_{j}$ the $(i,j)$ element $\mathrm{H}_{ij}^{'}$ leads to $\mathrm{H^{'F}}_{\alpha\beta}=\mathrm{R}_{ii}^{-1}\mathrm{H}_{ij}\mathrm{R}_{jj}$ where $\ket{\alpha}, \ket{\beta}$ are Fock basis satisfying $\bra{\alpha}\hat{c}_{i}^{\dagger}\hat{c}_{j}\ket{\beta}=1$.
Comparing this with the relation between non-Hermitian Hamiltonians $\mathrm{H^{F}}_{\alpha\beta}=\mathrm{H}_{ij}$, we get $\mathrm{H^{'F}}_{\alpha\beta}=\mathrm{R}_{ii}^{-1}\mathrm{H^{F}}_{\alpha\beta}\mathrm{R}_{jj}$.
In order to promote this relation into matrix multiplication, construct a diagonal matrix $\rm{R^{F}}$ in Fock space by the following relation
\begin{align}
\mathrm{R^{F}}_{\alpha\alpha}=\prod_{\{n_{i}\}}\mathrm{R}_{n_{i}n_{i}},
\end{align}
where Fock basis $\ket{\alpha}=\prod_{\{n_{i}\}}\hat{c}_{n_{i}}^{\dagger}\ket{0}$.
This enables us to write the relation as $\mathrm{H^{'F}}_{\alpha\beta}=\mathrm{R^{F}}_{\alpha\alpha}^{-1}\mathrm{H^{F}}_{\alpha\beta}\mathrm{R^{F}}_{\beta\beta}$.

Thus we have shown that
\begin{align}
\rm H^{'F}=(R^{F})^{-1}H^{F}R^{F}=H^{'F\dagger},
\end{align}
which satisfies $\rm (R^{F})^{-1}H^{F}R^{F}=R^{F\dagger}H^{F\dagger}((R^{F})^{-1})^{\dagger}$.
Natural consequence is that $\rm H^{F}=R^{F}R^{F\dagger}H^{F\dagger}(R^{F}R^{F\dagger})^{-1}$ which satisfies the necessary and sufficient condition for $\rm H^{F}$ to have a real energy eigenvalues~\cite{mostafazadeh2002pseudo1, mostafazadeh2002pseudo2, mostafazadeh2002pseudo3}.

\section{The reality condition of the ground state energy under PBC}

%By using many-body Hamiltonian matrix $\mathrm{H^F}$, the many-body ground state energy in PBC is real and thus physically stable.
%Here we define ground state $|\Psi^{G}_R\rangle$ whose real part of the energy eigenvalue is the smallest. 
%This can be explained by the following argument. 
%It is known that non-Hermitian matrix is pseudo-Hemitian if it satisfies
%\begin{align}
%\eta \mathrm{H} \eta^{-1} = \mathrm{H}^{\dagger}.
%\end{align}
%In our system, inversion symmetry $\hat{I}: c_{i}\rightarrow c_{L+1-i}$ safisfies the relation $I\mathrm{H}^{F}I^{-1}=\mathrm{H}^{F\dagger}.$
%Thus, the many-body Hamiltonian is pseudo-Hermitian.
%In general, a pseudo-Hermitian matrix has real eigenvalues or complex conjugate pairs~\cite{mostafazadeh2002pseudo1}.
%Thus, the uniqueness of the ground state is necessary and sufficient condition for the ground state energy eigenvalue to be real. 
%The pseudo-Hermiticity energy spectrum and the existence of the ground state are confirmed by the numerical calculations when $|\gamma|<|t_{1}|, |t_{2}|$.

The reality condition of the many-body ground-state energy can be shown by Bloch Hamiltonian. 
If we apply Bloch theorem, the $k$-space Hamiltonian $\rm{H}_{\rm{sin}}(k)$ is written as follows:
\begin{align}
\rm{H}_{\rm{sin}}(k)=\begin{pmatrix}
0 & t_1-\gamma+(t_2+\gamma)e^{ik} \\
t_1+\gamma+(t_2-\gamma)e^{-ik} & 0
\end{pmatrix},
\end{align}
where the energy eigenvalues are 
\begin{align}
E(k)^2&=t_{1}^2+t_{2}^2-2\gamma^2+(2t_1t_2+2\gamma^2)\cos{k}\nonumber\\&+2i\gamma(t_1+t_2)\sin{k}.
\end{align}
The Hamiltonian satisfies eigenvalue relation $E_{\rm occ}(k)=E_{\rm occ}^{*}(-k)$, where $E_{\rm occ}(k)$ is the energy eigenvalue with smaller real value. When the number of unit-cells is $N$, momentum $k$ is allowed for $k=\frac{2\pi n}{N}$, where $n=0, \cdot\cdot\cdot, n-1$.

The many-body ground state energy is expressed as
\begin{align}
&\sum_{n=0}^{n-1}E_{\rm occ}(2\pi n/N)\nonumber\\&=\sum_{n=0}^{n-1}- \Big\{ t_{1}^2+t_{2}^2-2\gamma^2+(2t_1t_2+2\gamma^2)\cos{\frac{2\pi n}{N}}\nonumber\\&+2i\gamma(t_1+t_2)\sin{\frac{2\pi n}{N}}\Big\}^{1/2}
\end{align}
For the odd number of $N$, the total energy of the ground-state is always real as we add all energy eigenvalues below the Fermi energy. This is because the energy eigenvalues are complex conjugate pairs with opposite momenta, and $E(0)=\pm(t_1+t_2)$ is real. On the other hand, when $N$ is even, the ground state energy is complex, since $k=\pi$ is allowed and $E(\pi)=\pm\sqrt{(t_1-t_2)^2-4\gamma^2}$ is imaginary for $t_2-2\gamma<t_1<t_2+2\gamma$.

Interestingly, the system is robust under perturbation that preserves generalized inversion symmetry; energy eigenvalue of the ground state remains real and our analysis can be carried out in the same way. Here, we give a simple proof showing that the ground state energy remains real even in the presence of a symmetry-preserving perturbation.
In the presence of generalized inversion symmetry $I$, the Hamiltonian satisfies the equation $IHI=H^{\dagger}$. 
The Hamiltonian is pseudo-Hermitian since it satisfies the equation
\begin{align}
\eta \mathrm{H} \eta^{-1} = \mathrm{H}^{\dagger}.
\end{align}
In general, a pseudo-Hermitian matrix has either real eigenvalues or complex conjugate pairs of eigenvalues~\cite{mostafazadeh2002pseudo1}.
If the symmetry-preserving perturbation is weak enough so that the unique ground state is maintained, the ground state energy remains real even when the perturbation term exists.

\section{Quantization of many-body polarization}
We have defined generalized polarization for non-Hermitian systems as:
\begin{align}
P^{LR}\equiv\lim_{N\rightarrow\infty}\frac{1}{2\pi}\text{Im}\ln\langle\Psi^{G}_{L}|e^{i\frac{2\pi}{N}\hat{X}}|\Psi^{G}_{R}\rangle \mod 1,
\end{align}
where $|\Psi^{G}_{R}\rangle(\ket{\Psi^{G}_{L}})$ is right (left) ground state. The polarization is quantized due to either $\hat{I}$or $\hat{S}$.

Under inversion symmetry, $I\hat{X}I^{-1}=-\hat{X}$ and $I|\Psi^{G}_R\rangle=|\Psi^{G}_L\rangle$ since $I\mathrm{H}I^{-1}I|\Psi^{G}_R\rangle=EI|\Psi^{G}_R\rangle=\mathrm{H}^{\dagger}I|\Psi^{G}_R\rangle$.
Thus, 
\begin{align} 
&\frac{1}{2\pi}\text{Im}\ln\langle\Psi^{G}_{L}|e^{i\frac{2\pi}{N}\hat{X}}|\Psi^{G}_{R}\rangle\nonumber\\
=&\frac{1}{2\pi}\text{Im}\ln\langle\Psi^{G}_{L}|I^{-1}Ie^{i\frac{2\pi}{N}\hat{X}}I^{-1}I|\Psi^{G}_{R}\rangle\nonumber\\
=&\frac{1}{2\pi}\text{Im}\ln\langle\Psi^{G}_{R}|e^{-i\frac{2\pi}{N}\hat{X}}|\Psi^{G}_{L}\rangle\nonumber\\
=&-\frac{1}{2\pi}\text{Im}\ln\langle\Psi^{G}_{L}|e^{i\frac{2\pi}{N}\hat{X}}|\Psi^{G}_{R}\rangle\mod 1,
\end{align}
which means $P^{LR}\equiv-P^{LR} \mod 1$. Therefore, $P^{LR}$ is quantized into $0$ or $1/2$.

Likewise, under chiral symmetry, $S|\Psi^{G}_R\rangle=|\Psi^{E}_R\rangle$,
because $S\mathrm{H}S^{-1}S|\Psi^{G}_R\rangle=ES|\Psi^{G}_R\rangle=-\mathrm{H}S|\Psi^{G}_R\rangle$, where $|\Psi^{E}_R\rangle$ is the most excited state.
Thus,
\begin{align} 
&\frac{1}{2\pi}\text{Im}\ln\langle\Psi^{G}_{L}|e^{i\frac{2\pi}{N}\hat{X}}|\Psi^{G}_{R}\rangle\nonumber\\
=&\frac{1}{2\pi}\text{Im}\ln\langle\Psi^{G}_{L}|S^{-1}Se^{i\frac{2\pi}{N}\hat{X}}S^{-1}S|\Psi^{G}_{R}\rangle\nonumber\\
=&\frac{1}{2\pi}\text{Im}\ln\langle\Psi^{E}_{L}|e^{i\frac{2\pi}{N}\hat{X}}|\Psi^{E}_{R}\rangle\mod 1,
\end{align}
which means that $P^{LR}_{\text{occ}}=P^{LR}_{\text{unocc}}$. By choosing appropriate unit-cell center such that $P^{LR}_{\text{occ}}+P^{LR}_{\text{unocc}}=0\mod 1$, we have quantization of $P^{LR}_{\text{occ}}$. 

\section{Critical point under PBC}
Here we prove why the many-body polarization changes at $t_1=t_2$ under PBC.
The many-body polarization is defined as 
\begin{align}
P^{LR}\equiv\lim_{N\rightarrow\infty}\frac{1}{2\pi}\text{Im}\ln\langle\Psi^{G}_{L}|e^{i\frac{2\pi}{N}\hat{X}}|\Psi^{G}_{R}\rangle \mod 1,
\end{align}
where $N$ is the number of unit-cells.
Within each unit-cell, there are two sublattice atomic positions: $x_1$ and $x_2$. Let us denote the positions of sublattice sites as $-0.5\leq x_1\leq x_2\leq0.5$, where the unit-cell center is located at $x=0$ and the length of the unit-cell is $1$. Since the system should respect the symmetries of the Hamiltonian, $x_1+x_2=0$.

Let us first consider the particular case for $x_2-x_1=0.5$.
In this case, reversing the role of $t_1$ and $t_2$ is equivalent to translating the system by a half lattice constant. Then $i$th site moves to the $(i+1)$-th site. 
Thus, the position operator becomes $\hat{X}\rightarrow \hat{X}+N/2 \mod{N}$ since $N$ electrons are translated by half of the lattice constant.
Thus, $P^{LR}$ of system with $t_1=\alpha, t_2=\beta$ is different by $1/2 \mod{1}$ from $P^{LR}$ with $t_2=\alpha, t_1=\beta$.

If $x_2-x_1\neq0.5$, on the other hand,  reversing the role of $t_1$ and $t_2$ is not equivalent to translating the system by half of the lattice constant.
However, when $N$ goes to infinity, the position operator $\hat{X}$ asymptotically becomes $\hat{X}+N/2 \mod{N}$ when $t_1$ and $t_2$ are reversed.
Since we have defined the polarization in $N\rightarrow\infty$ limit, the invariant is unchanged with arbitrary sublattice position choice that respects the symmetries.

%%%%%%%%%%%%%%%%%%%%%%%%%%%%%%%%%%%%%%%%%%%%%%%%%%%%%%%%%%%%%%%%%%%
\begin{figure}[b]
\centering
\includegraphics[width=8.5 cm]{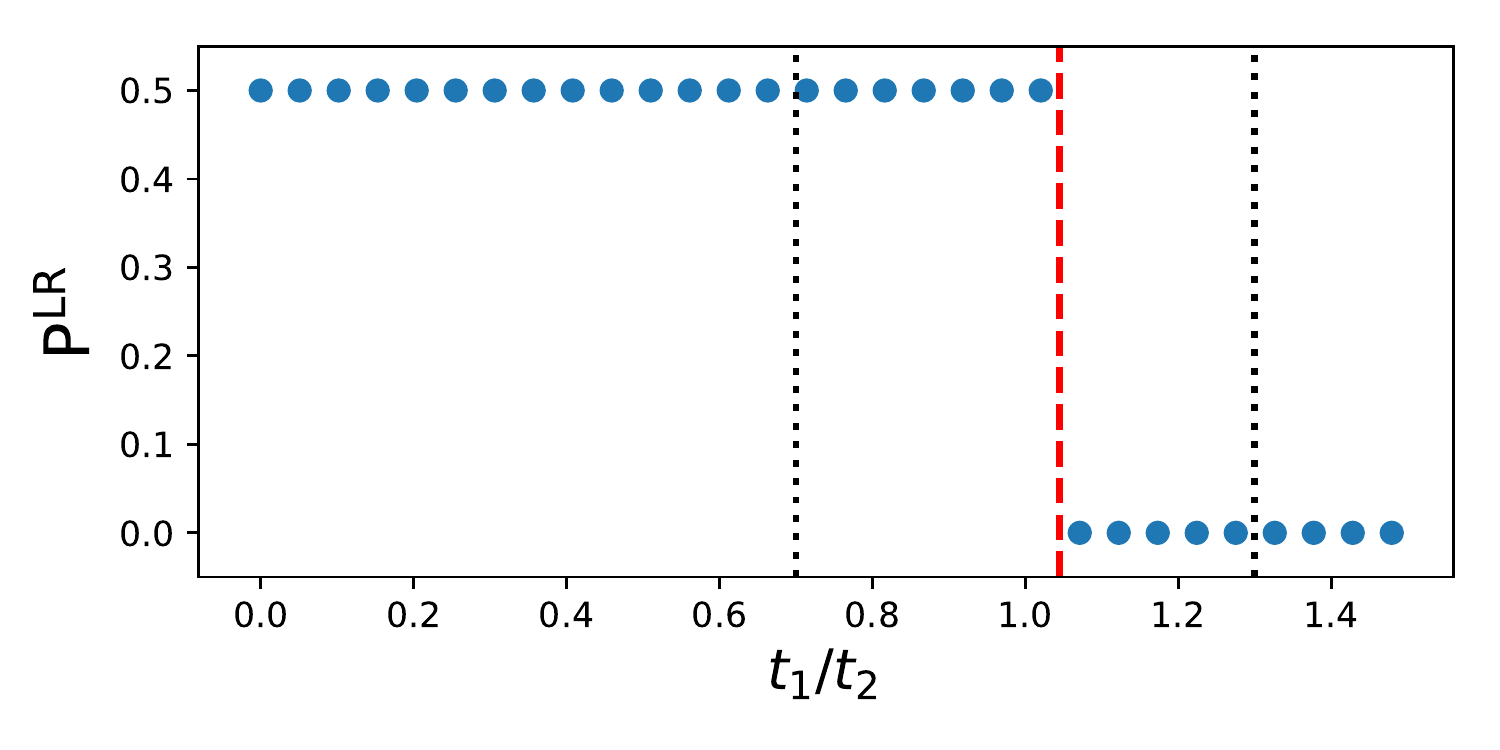}
\caption{
(Color Online) Non-Hermitian many-body polarization $P^{LR}$ as a function of $t_1$ with $t_2=1,~\gamma=0.3$. When $0<t_1<t_2$ ($t_1>t_2>0$), $P^{LR}=1/2$ ($P^{LR}=0$).
The critical point $t_1=\sqrt{t_2^2+\gamma^2}$ is consistent with the numerical study of the zero-modes and also previous studies under OBC. Thus, the bulk-boundary correspondence (BBC) can be successfully described by using $P^{LR}$.
The black dotted vertical lines indicate the locations of gap-closing points obtained by Bloch Hamiltonian, which cannot explain the BBC.
} \label{fig:sup2}
\end{figure}
%%%%%%%%%%%%%%%%%%%%%%%%%%%%%%%%%%%%%%%%%%%%%%%%%%%%%%%%%%%%%%%%%%%

\section{Confirmation in other model}\label{SM6}
The bulk-boundary correspondence is also confirmed by another type of SSH Hamiltonian introduced in Ref.~\onlinecite{yao2018edge}.
Here, the model Hamiltonian $\hat{H}=\hat{H}_0+\hat{H}_{\text{NH}}$:
\begin{align}
\hat{H}_{0}&=\sum_{i}\{(J-\Delta J(-1)^{i}\hat{c}_{i}^{\dagger}\hat{c}_{i+1}+h.c.)\}, \nonumber\\
\hat{H}_{\rm NH}&=\sum_{i}\gamma(\hat{c}_{2i}^{\dagger}\hat{c}_{2i-1}-\hat{c}_{2i-1}^{\dagger}\hat{c}_{2i}),
\end{align}
where $\hat{H}_{0}$ indicates the Hermitian SSH Hamiltonian with the intracell (intercell) hopping amplitude $t_{1}=J+\Delta J$ ($t_{2}=J-\Delta J$) while $\hat{H}_{\rm NH}$ denotes the non-Hermitian part describing asymmetric hopping processes.
Note that the asymmetric hopping interaction exists only for intracell electrons.

In this model, the critical point obtained in OBC is located at $t_1=\sqrt{t_2^2+\gamma^2}$ and agrees with the critical point of our bulk-invariant~\cite{yao2018edge} [See Fig.~\ref{fig:sup2}].
This critical point is different from the band gap closing point obtained from Bloch Hamiltonian.

%%%%%%%%%%%%%%%%%%%%%%%%%%%%%%%%%%%%%%%%%%%%%%%%%%%%%%%%%%%%%%%%%%%
\begin{figure}[b]
\centering
\includegraphics[width=8 cm]{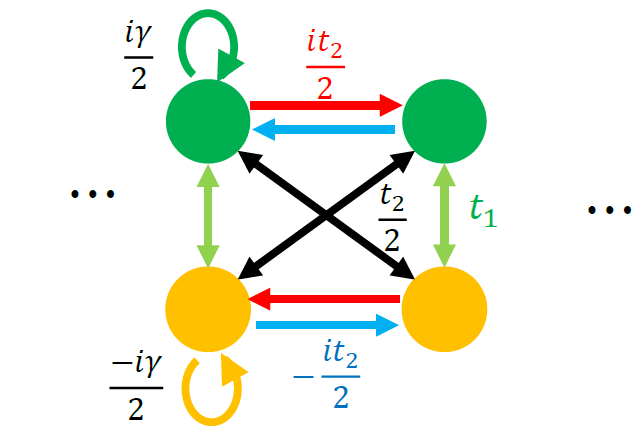}
\caption{
(Color Online) Physical setup that is equivalent to the 1D non-Hermitian system with asymmetric hopping.
This Creutz-ladder-like system is obtained by unitary basis transformation of the non-Hermitian Hamiltonian.
Notice that the asymmetric non-Hermitian interaction is now expressed as onsite gain and loss terms.
} \label{fig:sup3}
\end{figure}
%%%%%%%%%%%%%%%%%%%%%%%%%%%%%%%%%%%%%%%%%%%%%%%%%%%%%%%%%%%%%%%%%%%

The 1D non-Hermitian model with asymmetric hopping can be transformed into Creutz-ladder-like system with onsite gain and loss~\cite{yokomizo2019non}.
As written in Eq. S12, the asymmetric hopping interaction only exists in intra-cell and the $\hat{H}_{NH}$ is written as $i\gamma\sigma_y$ for each unit cell.
By the unitary transformation that changes unit cell basis: $\sigma_y\to-\sigma_z, \sigma_z\to\sigma_y$, one obtains a Creutz-ladder-like model with onsite gain and loss [See Fig.~\ref{fig:sup3}].
Such a system is experimentally realizable with ultracold fermionic atoms in optical lattice.

%\section{Failure of conventional edge entanglement entropy in non-Hermitian system}
%If we use conventional edge entanglement entropy $S_{2,\text{edge}}$ using only right eigenstate instead of  $S_{2,\text{edge}}^{LR}$ using left and right eigenstate, the entropy of the trivial phase is consistently zero. However, the entropy of the topological phase plummets as the asymmetric hopping amplitude increases. Thus it is clear that topological properties of the non-Hermitian system cannot be captured unless we use biorthogonal formalism.
%%%%%%%%%%%%%%%%%%%%%%%%%%%%%%%%%%%%%%%%%%%%%%%%%%%%%%%%%%%%%%%%%%%
%\begin{figure}[h]
%\centering
%\includegraphics[width=8.5 cm]{supfigure3_mod.png}
%\caption{
%(Color Online) The conventional edge entanglement entropy calculated using only right eigenstates. The number of sites is 14. The entropy of the trivial phases is consistently zero, whereas the topological phase's entropy plummets as the asymmetric hopping amplitude increases.
%}
%\end{figure}
%%%%%%%%%%%%%%%%%%%%%%%%%%%%%%%%%%%%%%%%%%%%%%%%%%%%%%%%%%%%%%%%%%%

%\bibliography{references}

\end{document}